\documentclass[12pt]{article}
 \input epsf.sty

\begin{document}

\def\ket{\rangle}
\def\bra{\langle}
\def\CA{{\cal A}}
\def\CB{{\cal B}}
\def\CC{{\cal C}}
\def\CD{{\cal D}}
\def\CE{{\cal E}}
\def\CF{{\cal F}}
\def\CG{{\cal G}}
\def\CH{{\cal H}}
\def\CI{{\cal I}}
\def\CJ{{\cal J}}
\def\CK{{\cal K}}
\def\CL{{\cal L}}
\def\CM{{\cal M}}
\def\CN{{\cal N}}
\def\CO{{\cal O}}
\def\CP{{\cal P}}
\def\CQ{{\cal Q}}
\def\CR{{\cal R}}
\def\CS{{\cal S}}
\def\CT{{\cal T}}
\def\CU{{\cal U}}
\def\CV{{\cal V}}
\def\CW{{\cal W}}
\def\CX{{\cal X}}
\def\CY{{\cal Y}}
\def\CZ{{\cal Z}}

\newcommand{\todo}[1]{{\em \small {#1}}\marginpar{$\Longleftarrow$}}
\newcommand{\labell}[1]{\label{#1}}
\newcommand{\bbibitem}[1]{\bibitem{#1}\marginpar{#1}}
\newcommand{\llabel}[1]{\label{#1}\marginpar{#1}}
\newcommand{\dslash}[0]{\slash{\hspace{-0.23cm}}\partial}

\newcommand{\sphere}[0]{{\rm S}^3}
\newcommand{\su}[0]{{\rm SU(2)}}
\newcommand{\so}[0]{{\rm SO(4)}}
\newcommand{\bK}[0]{{\bf K}}
\newcommand{\bL}[0]{{\bf L}}
\newcommand{\bR}[0]{{\bf R}}
\newcommand{\tK}[0]{\tilde{K}}
\newcommand{\tL}[0]{\bar{L}}
\newcommand{\tR}[0]{\tilde{R}}

\newcommand{\btzm}[0]{BTZ$_{\rm M}$}
\newcommand{\ads}[1]{{\rm AdS}_{#1}}
\newcommand{\ds}[1]{{\rm dS}_{#1}}
\newcommand{\eds}[1]{{\rm EdS}_{#1}}
\newcommand{\sph}[1]{{\rm S}^{#1}}
\newcommand{\gn}[0]{G_N}
\newcommand{\SL}[0]{{\rm SL}(2,R)}
\newcommand{\cosm}[0]{R}
\newcommand{\hdim}[0]{\bar{h}}
\newcommand{\bw}[0]{\bar{w}}
\newcommand{\bz}[0]{\bar{z}}
\newcommand{\be}{\begin{equation}}
\newcommand{\ee}{\end{equation}}
\newcommand{\bea}{\begin{eqnarray}}
\newcommand{\eea}{\end{eqnarray}}
\newcommand{\pat}{\partial}
\newcommand{\lp}{\lambda_+}
\newcommand{\bx}{ {\bf x}}
\newcommand{\bk}{{\bf k}}
\newcommand{\bb}{{\bf b}}
\newcommand{\BB}{{\bf B}}
\newcommand{\tp}{\tilde{\phi}}
\hyphenation{Min-kow-ski}

\def\apr{\alpha'}
\def\str{{str}}
\def\lstr{\ell_\str}
\def\gstr{g_\str}
\def\Mstr{M_\str}
\def\lpl{\ell_{pl}}
\def\Mpl{M_{pl}}
\def\varep{\varepsilon}
\def\del{\nabla}
\def\grad{\nabla}
\def\tr{\hbox{tr}}
\def\perp{\bot}
\def\half{\frac{1}{2}}
\def\p{\partial}
\def\perp{\bot}
\def\eps{\epsilon}

\renewcommand{\thepage}{\arabic{page}}
\setcounter{page}{1}

\rightline{hep-th/0204196}
\rightline{UPR-989-T}
\vskip 1cm
\centerline{\Large \bf Open Strings From ${\cal N} = 4$ Super Yang-Mills}
\vskip 0.5 cm
\renewcommand{\thefootnote}{\fnsymbol{footnote}}
\centerline{{\bf Vijay
Balasubramanian,\footnote{vijay@endive.hep.upenn.edu}
Min-xin Huang,\footnote{minxin@sas.upenn.edu}
Thomas S. Levi,\footnote{tslevi@student.physics.upenn.edu}
and
Asad Naqvi\footnote{naqvi@rutabaga.hep.upenn.edu}
}}
\vskip .5cm
\centerline{\it David Rittenhouse Laboratories, University of
Pennsylvania}
\centerline{\it Philadelphia, PA 19104, U.S.A.}

\setcounter{footnote}{0}
\renewcommand{\thefootnote}{\arabic{footnote}}

\begin{abstract}
Exploiting insights on strings moving in pp-wave backgrounds,
we show how open strings emerge from ${\cal N} = 4$ SU(N) Yang-Mills theory as fluctuations around certain states carrying R-charge of order $N$.   These states are dual to spherical D3-branes of $\ads{5} \times S^5$ and we reproduce the spectrum of small fluctuations of these states from Yang-Mills theory.
We discuss the emergence of the $G^2$ light degrees of freedom expected when $G$ such D3-branes nearly coincide.   The open strings running between the branes can be quantized easily in a Penrose limit of the spacetime.   Taking the corresponding large charge limit of the Yang-Mills theory, we reproduce the open string worldsheets and their spectra from field theory degrees of freedom.
\end{abstract}


\section{Introduction}
\label{intro}
The AdS/CFT correspondence states that the SU(N) Yang-Mills theory with 16 supersymmetries is equivalent to IIB string theory quantized around the $\ads{5} \times S^5$ background~\cite{jthroat}.   Recently, it has been shown that in some large charge limits of the field theory, corresponding to Penrose limits of spacetime\cite{blau}, the worldsheets of fundamental closed strings in spacetime can be explicitly reconstructed from Yang-Mills fluctuations~\cite{BMN}.    These results explicitly realize the long-standing goal of showing that Yang-Mills theories with many colours have a description in terms of closed string theories~\cite{Polyakov}.  The purpose of this paper is to show how open strings also emerge  from supersymmetric Yang-Mills theory, by quantizing fluctuations around states of large R-charge and displaying their worldsheets.\footnote{Recently~\cite{openstrings, parklee} showed how open strings emerge from gauge theories dual to string theories that have open strings in the perturbative spectrum.  In these cases, the dual field theory has quarks marking the endpoints of open string worldsheets.  Here we are interested in situations in which open strings emerge in pure supersymmetric Yang-Mills theory as fluctuations around states dual to D-branes in spacetime.   D-branes in pp-wave backgrounds and open strings propagating on them have been studied in~\cite{DP, kostasmarika, dbranes}.   In~\cite{bak} branes were found in the matrix model associated to pp-wave backgrounds. For earlier discussion of open strings in SU(N)
gauge theory, see \cite{ferrari}}

Open strings arise in string theory as the phonons of D-branes and so we seek the Yang-Mills descriptions of D-brane states of $\ads{5} \times S^5$.  Since most stable D-branes in AdS are infinite in size they are also infinitely massive, and so represent superselection sectors of the Yang-Mills.   However, $\ads{5} \times S^5$ also admits giant gravitons, namely finite-energy spherical D3-branes localized at the origin of AdS and propagating on the sphere~\cite{MST}.   If $G$ coincident D3-branes of the same size are present in the system, the open strings running between them should realize, at low energies, a $U(G)$ gauge theory on the $S^3 \times R$ worldvolume of the giants.

Giant graviton states of spacetime are created in Yang-Mills theory by determinant and sub-determniant operators as proposed in~\cite{BBNS} and confirmed in~\cite{CJR}.   In Sec.~2 we show how Yang-Mills theory  reproduces the spectrum of small fluctuations of giant gravitons.   We discuss the emergence of the $G^2$ degrees of freedom expected when $G$ giants nearly coincide.  In Sec.~3 we display a Penrose limit in which the open strings propagating on giants can be quantized simply.   Taking the corresponding large charge limit in Yang-Mills theory, we reconstruct the open string worldsheets from field theory degrees of freedom, and show that the field theory and string spectra match up to an $O(g_s)$ correction which we expect arises from open string one-loop effects in the pp-wave background.    In all these matches of spectra, energies in spacetime are mapped onto conformal dimensions of operators in the SU(N) Yang-Mills theory.   As we will see, the relevant operators are generically not BPS, but nevertheless their dimensions do not grow in the $N \rightarrow \infty$ limit.

Since we have a complete second-quantized formulation of ${\cal N} = 4$ Yang-Mills, this theory  is supposed to give us a non-perturbative description of strings.   If this is really so, various subsectors of the theory should contain the holographic duals to all possible string backgrounds.   How do we find these subsectors and the gravitational backgrounds that they describe?  Past experience with the M(atrix) model of M theory~\cite{BFSS}, the AdS/CFT correspondence~\cite{jthroat}, and pp-wave backgrounds~\cite{BMN} has suggested that large-charge, low-energy limits of string theory can give rise to holographic relations between field theories and string backgrounds.  Likewise, in Sec.~4 we argue that our results, coupled with the observations in~\cite{BN}, suggest that different large charge limits of Yang-Mills give rise in the infrared to theories dual to different backgrounds for strings.

\section{Spherical D3-branes and Their Fluctuations}

\paragraph{Scalar fluctuations:}
The best semiclassical description of a graviton with angular momentum
of order $N$  on the $\sph{5}$ of $\ads{5} \times \sph{5}$ is in terms of a large
D3-brane wrapping a 3-sphere and moving with some velocity~\cite{MST}.
This is the giant graviton.  The transition from a graviton mode in a
spherical harmonic to a macroscopic brane is explained by the Myers
effect~\cite{myerseffect} in the presence of flux on the $\sph{5}$.
In $\ads{5} \times \sph{5}$, the radius of the spherical D3-brane is
$\rho^2=lR^2/N$, where $l$ is the angular momentum on the $\sph{5}$ of the
state, $R$ is the radius of the sphere, and $N$ the total 5-form flux
through the 5-sphere. Since the radius of the D3 brane giant graviton
is bounded by the radius $R$ of the $\sph{5}$, there is an upper bound on
the angular momentum $ l \leq N$.

The spectrum of small fluctuations of the giant graviton was calculated in
\cite{DJM}. When the giant graviton expands into an $\sph{3}$ on the $\sph{5}$,
it has six transverse scalar fluctuations, of which four
correspond to fluctuations into  $\ads{5}$ and two are fluctuations within $\sph{5}$ . These vibration modes can be written as a superposition of  scalar spherical harmonics $Y_k$ on the unit  $S^3$.    In~\cite{DJM} it was found that the frequencies of
the four modes corresponding to fluctuations in $\ads{5}$ with wave-functions
$Y_k$ are given by
\begin{equation}
\omega_k=\frac{k+1}{R}
\label{frequency1}
\end{equation}
Similarly, the two vibration mode
frequencies corresponding to fluctuations in $\sph{5}$ are
\begin{equation}
\omega_k^-=\frac{k}{R},~~~~~\omega_k^+=\frac{k+2}{R}
\label{frequency2}
\end{equation}

\paragraph{Giants and their scalar fluctuations from CFT:}
In \cite{BBNS}, it was shown that  giant gravitons are dual
to states created by a family of subdeterminants:
\begin{equation}
O_l= {\rm subdet}{}_l Z \equiv \frac{1}{l!}\, \epsilon_{i_1 i_2 \cdots i_l a_1
a_2 \cdots
a_{N-l}}\,
\epsilon^{j_1 j_2 \cdots j_l a_1 a_2 \cdots
a_{N-l}}\,Z^{i_1}_{j_1}\, Z^{i_2}_{j_2}\,\cdots\, Z^{i_l}_{j_l}
\label{subdets}
\end{equation}
(So $O_{N}$ is the same as the
determinant of $\Phi$.)  Here, $Z=\phi^5+i \phi^6$ is a complex combination
of two of the six adjoint scalars in the $\CN=4$ theory.\footnote{The $\sph{5}$
in the bulk can be described by $X_1^2+\dots X_6^2=R^2$. The operator
$O_N$ in (\ref{subdets}) corresponds to a giant graviton moving in the $X^5,
X^6$ plane. The trajectory of such a giant will trace out a circle of radius
$(1-\frac{l}{N})R$ in this plane. Notice that the maximal giant with $l=N$ is
not really moving on the $\sph{5}$.   Its angular momentum arises from the Chern-Simons interaction on its worldvolume and the background flux.}
These subdeterminants have a bounded R-charge, with
the full determinant saturating the bound.  The bound on
the R-charge is the field theory explanation of the angular momentum
bound for giants.   A giant graviton is a 1/2 BPS state of the CFT and breaks the SO(6) R-symmetry of the $\CN = 4$ theory down to U(1) $\times$ SO(4).   The U(1) corresponds to the plane of motion of the giant gravitons while the SO(4) corresponds to the rotation group of the $\sph{3}$ worldvolume of the giants.  Under the U(1) $Z$ and $\bar{Z}$ have charges $\pm 1$ while the other scalars $\phi^i$ ($i = 1 \cdots 4$) of the Yang-Mills theory are neutral.   The giant gravitons in (\ref{subdets}) therefore carry a $U(1)$ charge $l$ and, being protected operators, their conformal dimensions are $\Delta = l$.  Under the SO(4), Z is neutral, but the $\phi^i$ transform as a {\bf 4}.

To map the fluctuations of a giant graviton to the CFT, we can replace $Z$ in (\ref{subdets}) by other operators, along lines similar to~\cite{BHK} for the dibaryon in the theory of D3-branes at a conifold singularity.  The resulting operator should carry the same U(1) charge $l$ as the giant.    Therefore, their conformal dimension in the free limit should take the   value $\Delta = l + \omega \, R$ where $\omega$ is appropriate fluctuation frequency in (\ref{frequency1}) or (\ref{frequency2}).\footnote{Recall that giant gravitons in global AdS map onto states of Yang-Mills theory on $\sph{3} \times R$ and that energy in spacetime maps to energy $E$ in the field theory.   Using the state-operator correspondence, the energy of states on $\sph{3} \times R$ maps to the dimension $\Delta = R \, E $ of operators on $R^4$, which we will typically discuss.}     Finally, since the scalar vibrations of giants are in  the $Y_k$ scalar spherical harmonics of $S^3$, i.e. the symmetric traceless representation of $SO(4)$, it is natural to use operators formed by the symmetric traceless
products of the four scalars $\phi^i$, ($i=1,2,3,4$).

Suppose $O^k$ is the $k$th symmetric traceless product of $\phi^i$.  Consider the operators:
\begin{equation}
 \label{mode3}
\CO_{m}^k =
\epsilon_{i_1\cdots i_la_1\cdots a_{n-l}}
\epsilon^{j_1\cdots j_la_1\cdots a_{n-l}}Z^{i_1}_{j_1}\cdots
Z^{i_{l-1}}_{j_{l-1}}(D_m Z \, O^k)^{i_{l}}_{j_{l}}
\end{equation}
$\CO^k_m$ are operators with U(1) charge $l$,  in the $k$th symmetric traceless representation of SO(4), and have dimension $\Delta = l+k + 1$.   The index $m=1 \cdots 4$ refers to the four Cartesian directions of $R^4$ in radial quantization of $\sph{3} \times R$.  Clearly, $\CO^k_m$ has the quantum numbers to match the AdS polarized fluctuations with spectrum (\ref{frequency1}).   (Note that unlike (\ref{subdets}) we have not normalized these operators to have unit two-point functions.)

Now consider
\begin{eqnarray}
\label{mode1}
\CO^k_- &=&
\epsilon_{i_1\cdots i_li_{l+1}a_1\cdots a_{n-l-1}}
\epsilon^{j_1\cdots j_lj_{l+1}a_1\cdots
a_{N-l-1}}Z^{i_1}_{j_1}\cdots
Z^{i_l}_{j_l}(O^k)^{i_{l+1}}_{j_{l+1}}
\\
\label{mode2}
\CO^k_+ &=&
\epsilon_{i_1\cdots i_{l+3}a_1\cdots a_{n-l-3}}
\epsilon^{j_1\cdots j_{l+3}a_1\cdots a_{n-l-3}}Z^{i_1}_{j_1}\cdots
Z^{i_l}_{j_l}Z^{i_{l+1}}_{j_{l+1}}\overline{Z}^{i_{l+2}}_{j_{l+2}}(O^k)^{i_{l+3}}_{j_{l+3}}
\end{eqnarray}
These operators have U(1) charge $l$, are in $k$th symmetric traceless representation of SO(4), and have conformal dimensions $\Delta^- = l + k$ and $\Delta^+= l + k + 2$.   Clearly we have found operators with quantum numbers matching the $\sph{5}$ polarized fluctuations whose spectrum is (\ref{frequency2}).   (Again, we have not chosen to normalize these operators to have unit two-point functions.) Note that the operators (\ref{mode1},\ref{mode2}) cannot be constructed for the maximal giant graviton, i.e., when $l=N$.   The corresponding analysis of fluctuations in~\cite{DJM} leads to a similar conclusion since the relevant equations are ill-defined for the maximal giant.

In general, most fluctuations of giant gravitons are not
BPS~\cite{DJM} and so we expect anomalous dimensions to develop
quantum mechanically.  From the spacetime point of view these
would be studied by finding solutions to the open-string loop
corrected equations of motion of a D-brane.   (The DBI action used
in~\cite{DJM} included all $\alpha^\prime$ corrections at disk
order but not string loop corrections.)   Since these corrections
are hard to compute in spacetime it is interesting to examine them
in the field theory.  In the appendix we show that the
interactions of $\CN = 4$ Yang-Mills do produce anomalous
dimensions for (\ref{mode3},\ref{mode1},\ref{mode2}), but,
surprisingly, these corrections do not grow with $N$.  As shown in
the appendix the anomalous dimension is $(J-1)g_s/\pi$.     At
weak coupling and large $N$, therefore, these are non-BPS
operators whose dimensions are protected from large corrections.

\paragraph{Multiple giants from Yang-Mills:}
Consider a CFT states made by the product of $G$ identical giant graviton operators.  This should represent $G$ giants of the same size moving in concert on the $\sph{5}$.  Such a group of giants should have $G^2$ strings stretching between them.   At low energies these string should give rise to a $U(G)$ gauge theory living on the worldvolume of the spherical D3-branes.   The spectra that we described above we derived from fluctuations of a single giant, and therefore apply to the $U(1)^G$ part of this low-energy gauge theory.   Below we will display candidate operators dual to the expected $6 G^2$ scalar fluctuations of the branes.  In the quadratic limit relevant to small fluctuations all of these will have the same spectrum as we see below.  (Again, there are small quantum corrections to the spectrum that are negligible in the large N limit.)

For simplicity consider two maximal giant gravitons, corresponding to
the product of two determinants in the CFT $\det_1Z \, \det_2Z$.  Here we have introduced labels analogous to Chan-Paton indices for each of the determinants representing a giant graviton. Taking $x_1$ and $x_2$ to be coordinates on the $\sph{3}$ on which the Yang-Mills theory is defined, we could define the operator $\det Z(x_1) \, \det Z(x_2)$ so that it makes sense to treat them as distinguishable in this way.   After constructing the operators of interest to us as described in the text we later let $x_1 \rightarrow x_2$.    For each of the operators $\CO^k_{m,+,-}$ describing scalar fluctuations on a single giant we might expect four operators here.  Two of these would correspond to separate vibrations of each of the two giants and should be given by $\CO^k_{11;m,+,-} \equiv \CO^k_{1;m,+,-} \, \det_2 Z$ and  $\CO^k_{22;m,+,1} \equiv \det_1 Z \, \CO^k_{2;m,+,-}$ where $\CO^k_{i;m,+,-}$ is a fluctuation on giant $i$ as described above. Two of the fluctuations would arise from off-diagonal components of the U(2) matrices, which in turn arise from strings going between the two branes.   It is natural that these vibrations should arise from operators that intertwine the gauge indices in the two determinants:
\begin{eqnarray}
\CO^k_{12;-} &=&
\epsilon_{i_1\cdots i_N} \epsilon^{j_1\cdots
j_N}Z^{i_1}_{j_1}\cdots Z^{i_{N-1}}_{j_{N-1}}(O^{k})^{i_N}_{l_N} \, \, \,
\epsilon_{k_1\cdots k_N} \epsilon^{l_1\cdots
l_N}Z^{k_1}_{l_1}\cdots Z^{k_N}_{j_N}
\label{inter1}
\\
\CO^k_{21;-} &=&
\epsilon_{i_1\cdots i_N} \epsilon^{j_1\cdots
j_N}Z^{i_1}_{j_1}\cdots Z^{i_{N}}_{l_{N}}
\, \, \,
\epsilon_{k_1\cdots k_N}
\epsilon^{l_1\cdots l_N}Z^{k_1}_{l_1}\cdots
Z^{k_{N-1}}_{l_{N-1}}(O^k)^{k_N}_{j_N}
\label{inter2}
\end{eqnarray}
Here the $O^k$ are the symmetric traceless products of scalars mentioned earlier.  The other operators $\CO^k_{ij;+,m}$ are written similarly.  The indices that we introduced on the two determinants have played the  role of Chan-Paton factors of open strings stretched between two giants.  It appears that the four operators we find have a natural interpretation as the expected operators from the adjoint representation of $U(2)$ group.   However, there is a subtlety.  When $x_1 \rightarrow x_2$, $O_{12,-}^k = O_{21,-}^k$ and  $O_{11,-}^k = O_{22,-}^k$ by an exchange of dummy indices.  What is more, it can be shown that in this limit the operators $O_{11,-}^k = N \, O_{12,-}^k$ also.  This is surprising at first because $O_{12}$ intertwines indices between two determinants, but this fact can be shown as follows.   First, observe that (\ref{inter1}) is zero when $l_N \neq j_N$.  In this case there exists an $l_x = j_n$ where $1 \leq x \leq N-1$, and so the sum overs the permutations of $k_1,\cdots k_N$ gives zero.  We are left with $l_N = j_N$ in which case the operator is unchanged by switching these two indices.  Thus $O_{11;-}^{k}$ is proportional to $O_{12;-}^{k}$. The proportionality factor between these operators is $N$ because the former has
$N^2$ choices of $j_N, l_N$,  while requiring $l_N=j_N$ leaves $N$ choices.

In fact this is exactly what we should expect since coincident D-branes are identical and there is no difference between the strings running between different pairs of branes; the change of dummy variables relating (\ref{inter1}) and (\ref{inter2}) when $x_1 \rightarrow x_2$ is an example of this.   To display the four strings running between two branes we have to separate the branes from each each other.     The generalization of this discussion to $G$ giants and the associateed $G^2$ degrees of freedom from the adjoint of $U(G)$ is obvious.

There is some issue as to whether multiple less-than-maximal giant gravitons are described by a product of subdeterminant operators or by a sudeterminant of products (see~\cite{BBNS,CJR,BHK}).   Above we restricted ourself to the largest giants for which this issue does not arise since the determinant of products is the product of determinants.    It would be interesting to test our proposal that fluctuations of strings between branes are described by intertwined operators, by taking one of them to be a maximal giant and another to be a smaller one.    Strings running between such branes are stretched and should have a corresponding gap in their spectra.  It would be interesting to test this by trying to match the vibrational energies of strings stretched between the maximal and next-to-maximal sized giants.

\section{Open string world sheets from CFT}
\subsection{Penrose limits for open strings}
We will see that for open strings moving with a large angular
momentum on the giant graviton D3 brane, we can construct the open
string world sheet  in the $\CN=4$ $SU(N)$ Yang-Mills theory. To
that end, we start by looking at the geometry seen by such an open
string.   The $\ads{5} \times \sph{5}$ metric is:
\begin{eqnarray*}
ds^2&=&R^2[ -dt^2 \cosh^2 \rho + d \rho^2 + \sinh^2 \rho ~ d\Omega_3^2+d \psi^2 \cos^2 \theta + d\theta^2 + \sin^2 \theta d {\Omega_3^\prime}^2] \\
d{\Omega_3^\prime}^2& =& d \varphi^2 + \cos^2 \varphi d \eta^2 + \sin^2 \varphi d \xi^2
\end{eqnarray*}
where $R = (4\pi \alpha^{\prime2} g N)^{1/4}$ is the AdS scale.
Consider the near maximal giant graviton at $\theta \sim \pi/2$
which is moving in the $\psi$ direction.\footnote{Strictly
speaking, the maximal giant is at rest and all its angular
momentum comes from the five-form flux.}  The world volume of the
giant spans $(t,\varphi, \eta, \xi)$ and the giant graviton is at
$\rho=0$ and $\theta={\pi \over 2}$. We want to find the geometry
seen by an open string ending on the giant graviton, moving rapidly
in the $ \eta $ direction at the equator of $\sph{3}$ given
by $\varphi=0$. We define light cone coordinates
${\tilde{x}}^{\pm}={ {t\pm\eta} \over 2}$, a new coordinate $\chi
= {\pi \over 2}- \theta$, and focus on the region near
$\rho=\chi=\varphi=0$ by rescaling:
\begin{equation}
x^+ = \tilde{x}^+, ~~~ x^-=R^2 \tilde{x}^-, ~~~ \rho={r \over R}, ~~~ \chi={y \over R}, ~~~
\varphi={u \over R}, ~~~R \rightarrow \infty .
\end{equation}
In this limit, the metric becomes,
\begin{equation}
ds^2=-4 dx^+ dx^--(\vec{r}^2+\vec{y}^2+\vec{u}^2)(dx^+)^2+d\vec{y}^2+d\vec{u}^2+d\vec{r}^2\
\label{pp}
\end{equation}
where $\vec{u}$ and $\vec{y}$ parameterize points on two $R^2$s and
$\vec{r}$ parameterizes points on $R^4$.   The 5-form flux that supports the $\ads{5} \times \sph{5}$ background becomes
\begin{equation}
F_{+1234} = F_{+5678} = {\rm Const} \times \mu
\end{equation}
in this limit, thereby breaking the $SO(8)$ isometry of the metric (\ref{pp}) to SO(4) $\times$ SO(4).   We find that the open string sees the standard pp wave geometry. The light cone action
becomes \cite{metsaev},
\begin{equation}
S=\frac{1}{4\pi \alpha'} \int ~dt \int^{2\pi \alpha' p^+}_0
d \sigma \Bigl[  {\partial_\tau X^I}{\partial_\tau X_I}  - {\partial_\sigma X^I}{\partial_\sigma X_I} - \mu^2 \, X^I X_I  + 2i \bar{S} (\dslash+\mu \Pi)S \Bigr]
\label{action}
\end{equation}
where $\Pi=\Gamma^{1234}$ and $S$ is a Majorana spinor on the worldsheet
and a positive chirality spinor ${\bf 8_s}$  under SO(8) which is the group
of rotations in the eight transverse directions. The $X^I$ tranform as ${\bf
8_v}$ under this group.    The fermionic term in the action breaks the SO(8) symmetry that is otherwise present to SO(4) $\times$ SO(4).
The open strings we are interested in have Neumann boundary
conditions in the light cone directions $x^{\pm}$ and $\vec{u}$
and Dirichlet boundary conditions in the six transverse directions
parameterized by
$\vec{r}$ and $\vec{y}$.
\begin{equation}
\partial_\sigma {X^\alpha}
=\partial_\tau {X^i}  =  0
\label{bcond}
\end{equation}
where $\alpha=7,8$. The coordinates used in (\ref{pp}), $\vec{u}=(x^7,x^8)$,
$\vec{r}=(x^1,x^2,x^3,x^4)$ and $\vec{y}=(x^5,x^6)$.
Such open strings were
quantized by Dabholkar and Parvizi in \cite{DP}. Here, we quote their
results.
The spectrum of the light cone Hamiltonian ($H \equiv -p_+$) is
\begin{eqnarray}
H& = & E_0+E_\CN \nonumber \\
E_0&=&\mu \Bigl( \sum_{\alpha=7,8} \bar{a}_0^\alpha a_0^\alpha -2 i S_0 \Gamma^{56}
S_0 +e_0 \Bigr) \nonumber \\
 E_\CN& = &  \Bigl({1 \over 2} \sum_{n \neq 0} \omega_n~a_n^i a_{-n}^i + i \sum_{n \neq 0}
\omega_n ~S_n S_{-n} \Bigr).  \label{lightham}
\end{eqnarray}
where we have defined the bosonic and fermionic creation and annihilation operators $a_n^i$
and $S_n$  as in~\cite{DP}.  Here $e_0=1$ is the zero point energy for the D3 brane and
\begin{equation}
\omega_n=\mathrm{sign}(n)\sqrt{\Bigl({n \over {2 \alpha^\prime p^+}}\Bigr)^2+\mu^2} \, .
\label{freq}
\end{equation}
There are only two bosonic zero modes coming from two directions in light cone gauge
which have Neumann boundary conditions (these would have been momentum
modes but in the pp wave background, the zero mode is also a harmonic
oscillator). The fermionic zero mode is $S_0$ which transforms in ${\bf 8_s}$
of SO(8).

The D3 brane occupies
$x^+,x^-, x^7,x^8$ and has six transverse coordinates $x^1 \cdots x^6$.
In the light cone, only an SO(2)${}_U$ subgroup of the SO(1,3) symmetry of the
D3 brane world volume is visible. In addition, the  SO(6) group transverse to the D3-brane
is broken down to SO(2)${}_Z \times $ SO(4)  by the 5-form background flux.   Hence
we have the embedding
\begin{equation}
SO(8) \supset SO(2)_Z \times SO(2)_U  \times SO(4)
\labell{symm1}
\end{equation}
The spinor ${\bf 8_s}$ decomposes as
\begin{equation}
{\bf 8_s} \rightarrow ({\bf 2},{\bf 1})^{(\half,\half)}\oplus ({\bf \bar{ 2}},{\bf 1})^{(-\half,-\half)}
\oplus ({\bf 1},{\bf 2})^{(\half,-\half)}\oplus ({\bf 1},{\bf \bar{2}})^{(-\half,\half)}
\end{equation}
where the superscripts denote SO(2)${}_Z \times$ SO(2)${}_U$, and
we have written SO(4) representations as representations of SU(2)
$\times$ SU(2). The fermionic zero modes $S_0$ can be arranged
into fermionic creation and annihilation operators:
\begin{eqnarray}
\bar{\lambda}_\alpha  \equiv S_{0\alpha}^{(\half,\half)} &,&
{\lambda}_\alpha  \equiv S_{0\alpha}^{(-\half,-\half)}  \nonumber \\
\bar{\lambda}_{\dot{\alpha}}  \equiv S_{0{\dot{\alpha}}}^{(-\half,\half)} &,&
{\lambda}_{\dot{\alpha}}  \equiv S_{0{\dot{\alpha}}}^{(\half,-\half)}
\label{lambdas}
\end{eqnarray}
The commutation relations are
\begin{equation}
\{\bar{\lambda}_\alpha,\lambda^\beta \}= \delta^\beta_\alpha ~~~~,~~~~~
\{{\lambda}_{\dot{\alpha}},{\bar{\lambda}}^{\dot{\beta}} \}= \delta^{\dot{\beta}}_{\dot{\alpha}}.
\end{equation}
The energy contribution from the zero mode oscillators is given by
\begin{eqnarray}
E_0&=& m(\bar{a}_0^7a_0^7+\bar{a}_0^8a_0^8+\bar{\lambda}_\alpha \lambda^\alpha
-\bar{\lambda}_{\dot{\alpha}}\lambda^{\dot{\alpha}} + 1)\\
& = &  m(\bar{a}_0^7a_0^7+\bar{a}_0^8a_0^8+\bar{\lambda}_\alpha \lambda^\alpha
+{\lambda}_{\dot{\alpha}}\bar{\lambda}^{\dot{\alpha}} -1)
\end{eqnarray}
as in~\cite{DP}.   We choose $\bar{\lambda}_\alpha$ and $\lambda_{\dot{\alpha}}$ as creation
operators.\footnote{This convention differs from~\cite{DP} but is convenient for us.}   The vacuum state  is invariant under SO(4) $\times$ SO(2)${}_U$
and carries charge $-1$ under SO(2)${}_Z$:
\begin{equation}
a_0^r| - 1 , 0 \ket = 0 , ~~~~\lambda^\alpha | -1,0 \ket = 0, ~~~~ {\bar {\lambda}}^{\dot{\alpha}}
|-1, 0 \ket = 0
\label{grounddef}
\end{equation}
Other modes with zero worldsheet momentum ($n=0$) are contructed by acting with creation
operators $\bar{\lambda}_\alpha$, $\lambda_{\dot{\alpha}}$ and $\bar{a}_0^{r}$ . The vacuum
state $| -1,0 \ket$ has $E_0=-1$ and carries $-1$ units of angular momentum
in the $x^5 x^6$ direction. Since the maximal giant graviton that we are considering carries
angular momentum $N$ in this direction, we see that  $N-1$ is the total angular momentum of the giant and the ground state of its open strings in our Penrose limit.  Likewise the fermionic contribution to the energy ground state lowers it to $N-1$ from the value $N$ for the maximal giant.  Below we will present the complete perturbative spectrum of the string quantized in this way and map all the states to operators of $\CN =4$ Yang-Mills theory.

\subsection{Open string world sheet in SYM}
String theory in global $\ads{5} \times \sph{5}$ is dual to $\CN = 4$ Yang-Mills theory on $\sph{3} \times R$.   States in spacetime map to states of the field theory, and by the state-operator correspondence for conformal theories, are related to operators on $R^4$.    The global symmetry of the theory is SO(4) $\times$ SO(6) where SO(4) is the rotation group of $R^4$ corresponding to the SO(4) appearing in (\ref{symm1}).   SO(6) is the R-symmetry group, corresponding to the rotation group of $\sph{5}$ in the bulk spacetime.  The Yang-Mills theory has six adjoint scalar fields $\phi^1 \cdots \phi^6$ which transform as the fundamental of SO(6).   The  complex combinations $Z=\phi^5+i \phi^6$, $Y=\phi^3+i \phi^4$, $U=\phi^1 + i\phi^2$  are charged under three different SO(2) subgroups of SO(6), SO(2)$_{Z,Y,U}$, which correspond to rotations in three independent planes of the bulk $\sph{5}$.   We will denote charges under these SO(2) groups as $J_{Z,Y,U}$.   As we have discussed, giant gravitons carry a charge of order $N$ under SO(2)$_Z$, and are created by subdeterminant operators.  The Penrose limit of open strings on giants corresponds to strings moving in the spacetime direction corresponding to $Y$.  We will propose a field theory description of the worldsheets of open strings on the maximal giant graviton.

We will denote the conformal dimenion of the operators dual to such strings by
\begin{equation}
{\tilde\Delta} = N + \Delta
\end{equation}
where the additive $N$ arises because the background giant has this dimension. Mapping the data of the Penrose limit to the field theory we find that the conformal dimension $\Delta$ of the excitation above the giant and $SO(2)_Y$ charges of these states are related to the lightcone Hamiltonian (\ref{lightham}) of strings as~\cite{BMN}
\begin{equation}
H = -p_+ = 2p^- = \Delta - J_Y \, = {\tilde\Delta} - N - J_Y.
\end{equation}
We are going to consider states of fixed $p_+$ so that $\Delta \approx J_Y$. Likewise the other lightcone momentum maps as
\begin{equation}
-p_- = 2p^+ = {\Delta + J_Y \over R^2} \, ,
\end{equation}
where $R$ is the AdS scale.   Note that to have a fixed non-zero value of $p_-$, $J_Y$ must be of order $\sqrt{N}$.  The contribution to Hamiltonian from higher oscillator modes of the lightcone string (\ref{freq}) then translates into \footnote{We have set $\mu =1$ to make the comparison
with field theory.}
\begin{equation}
( \tilde\Delta - N - J_Y)_n = (\Delta - J_Y)_n = \omega_n = \sqrt{1 + {\pi g N n^2 \over J_Y^2}}
\label{spectrum}
\end{equation}
in the Yang-Mills theory up to small corrections that vanish at large $N$.

\paragraph{The ground state: }   From the previous section, the ground state of strings on maximal giants carries SO(2)$_Z$ charge $-1$ and lightcone energy $-1$.  So the overall state including the giant   carries an SO(2)$_Z$ charge $N - 1$ and $\Delta - J_Y = -1$.   Furthermore, we achieve the Penrose limit by considering states with SO(2)$_Y$ charge $J$ of order $\sqrt{N}$.   To describe the ground state of the string we therefore seek an operator that is a modification of the $\det Z$ creating the maximal giant which has the charges just listed. A suitable candidate is:
\begin{equation}
\epsilon_{i_1\cdots i_N} \epsilon^{j_1\cdots
j_N}Z^{i_1}_{j_1}\cdots Z^{i_{N-1}}_{j_{N-1}}
(Y Y \cdots Y)^{i_N}_{j_N} ~~~~
\leftrightarrow ~~~~
| G_N ; -1, 0 \ket
\label{opdef}
\end{equation}
where we have inserted a product of $J$ Ys in place of one Z. (We
have chosen not to normalize this operator to have a unit
two-point function.)  The notation $ |G_N ; -1, 0 \ket$ indicates
a single maximal giant graviton with an open string in its ground
state as defined in (\ref{grounddef}).  The Zs create the D3 brane
giant graviton and, as we show below, the string of Ys explicitly
reconstructs the worldsheet of open string propagating on a giant
in a Penrose limit.The absence of a trace on the indices of the
product of Ys will be responsible for making this an open string
worldsheet, and in the presence of multiple giants, Chan-Paton
factors will emerge from the ability to intertwine these indices
between different giant operators.

The operator (\ref{opdef}), just like the small fluctuations
(\ref{mode3},\ref{mode1},\ref{mode2}), is not BPS and therefore
receives quantum corrections to its dimension.  Surprisingly, as
we show in the Appendix, these corrections do not grow with $N$
and lead to an anomalous dimension of $(J-1)g_s/\pi$.  This extra
piece is very small compared to the BMN anomalous dimension
$\frac{g_sN}{J^2}$  if we take $g_2=\frac{J^2}{N}$ small to
suppress the non-planar contributions anomalous dimensions
\cite{nonplanar, Constable}.   Nevertheless, the anomalous
dimension of our operator differs from the prediction from the
spacetime calculation of Dabholkar and Parvizi \cite{DP} by
$(J-1)g_s/\pi$. This difference, being of $O(g_s)$, suggests a
loop open-string effect. Perhaps the interesting phenomenon of a
non-BPS operator acquiring only a small anomalous dimension occurs
because of the restoration of supersymmetries in the pp-wave limit
\cite{susys}.

\paragraph{Rest of the zero modes: } The remainder of the zero modes on the string worldsheet arose from the ground state (\ref{grounddef}) by the action of the creation operators $\bar{\lambda}_\alpha$, $\lambda_{\dot{\alpha}}$ and $\bar{a}_0^r$.   In Yang-Mills theory we can construct operators corresponding to these states by inserting into the string of the Ys the two scalars $\phi^{1,2}$ and four of the 16 gaugino components of $\CN =4$  Yang-Mills which have SO(2)$_Y$ charge $J_Y = 1/2$ and SO$_Z$ charge $J=1/2$.  These gaugino components in which we are interested transform as $({\bf 2,1 })$ and $({\bf 1, 2})$ under the SO(4) = SU(2) $\times$ SU(2) rotation group of four dimensional Yang-Mills and so we will collect into two spinors $\psi_\alpha$ and $\psi_{\dot\alpha}$.   This matches the charges carried by the four creation operators $\bar\lambda_\alpha$ and $\lambda_{\dot\alpha}$ identified on the lightcone string worldsheet in (\ref{lambdas}).    So we identify the string worldsheet zero mode creation operators with operator insertions into (\ref{opdef}) as follows:
\begin{eqnarray}
\bar{a}_0^{7,8}  ~~~~ &\leftrightarrow& ~~~~ \phi^{1,2} \nonumber
\\
\bar\lambda_\alpha , \lambda_{\dot\alpha}
~~~~
&\leftrightarrow&
~~~~
\psi_\alpha, \psi_{\dot\alpha}
\label{corresp1}
\end{eqnarray}
For example,
\begin{equation}
\bar\lambda_\alpha |G_N ; -1,0\ket
~~~~
\leftrightarrow
~~~~
\epsilon_{i_1\cdots i_N} \epsilon^{j_1\cdots
j_N}Z^{i_1}_{j_1}\cdots Z^{i_{N-1}}_{j_{N-1}}
\left( \sum_{l=0}^{J_Y} \, Y^l \, \psi_\alpha Y^{J_Y - l} \right)^{i_N}_{j_N}
\end{equation}
Each action of a zero mode operator on the lightcone string vacuum adds a similar sum to the dual field theory operator.  It is interesting to see a detailed match between  field theory operators and the quantum numbers for states created by acting by worldsheet fermionic zero modes.   Each of these operators takes the form
$\frac{1}{N!}\epsilon_{i_1\cdots i_N} \epsilon^{j_1\cdots
j_N}Z^{i_1}_{j_1}\cdots Z^{i_{N-1}}_{j_{N-1}} \,
\CV^{i_N}_{j_N}$ with $\CV$ given as below:
\begin{center}
\begin{tabular}{|r|l|c|l|}
\hline
State & Rep. & $H = \Delta - J_Y$ & $\CV$ \\
\hline
$|-1,0\ket$ & $({\bf 1},{\bf 1})^{(-1,0)}$ & $- 1$& $Y^{J_Y}$\\
$\bar\lambda_\alpha| -1,0\ket$&
$({\bf 2},{\bf 1})^{(-\half,\half)}$ & $0$
& $\sum_{l=0}^{J_Y} \, Y^l \, \psi_\alpha Y^{J_Y - l} $
\\
$\lambda_{\dot{\alpha}}|-1,0\ket$ &$({\bf 1},{\bf 2})^{(-\half,-\half)}$ &0 &$
 \sum_{l=0}^{J_Y} \, Y^l \, \psi_{\dot{\alpha}} Y^{J_Y - l} $ \\
$\bar\lambda_\alpha\bar\lambda_\beta| -1,0\ket$&
$({\bf 1},{\bf 1})^{(0,1)}$ & $1$
& $ \sum_{l_1,l_2=0}^{J_Y} \, Y^{l_1} \, \psi_\alpha Y^{l_2}\psi_{\beta}Y^{J_Y-l_1-l_2} $\\
$\bar\lambda_\alpha \lambda_{\dot{\alpha}}| -1,0\ket$&
$({\bf 2},{\bf 2})^{(0,0)}$ & $1$
& $ \sum_{l_1,l_2=0}^{J_Y} \, Y^{l_1} \, \psi_\alpha Y^{l_2}\psi_{\dot{\alpha}}Y^{J_Y-l_1-l_2} $\\
 $\lambda_{\dot{\alpha}}\lambda_{\dot{\beta}}| -1,0\ket$&
$({\bf 1},{\bf 1})^{(-1,-1)}$ & $1$
& $ \sum_{l_1,l_2=0}^{J_Y} \, Y^{l_1} \, \psi_{\dot{\alpha}} Y^{l_2}\psi_{\dot{\beta}}Y^{J_Y-l_1-l_2} $\\
$\bar\lambda_\alpha\lambda_{\dot{\alpha}}\lambda_{\dot{\beta}}| -1,0\ket$&
$({\bf 2},{\bf 1})^{(-\half,-\half)}$ & $2$
& $ \sum_{l_i=0}^{J_Y} \, Y^{l_1} \,\psi_\alpha  Y^{l_2}\ \psi_{\dot{\alpha}} Y^{l_3} \psi_{\dot{\beta}}Y^{J_Y-\sum {l_i}} $\\
$\bar\lambda_\alpha\bar\lambda_{{\beta}}\lambda_{\dot{\beta}}| -1,0\ket$&
$({\bf 1},{\bf 2})^{(\half,\half)}$ & $2$
& $ \sum_{l_i=0}^{J_Y} \, Y^{l_1} \,\psi_\alpha  Y^{l_2}\ \psi_{{\beta}} Y^{l_3} \psi_{\dot{\beta}}Y^{J_Y-\sum l_i} $\\
$\bar\lambda_\alpha\bar\lambda_{{\beta}}\lambda_{\dot{\alpha}}\lambda_{\dot{\beta}}| -1,0\ket$&
$({\bf 1},{\bf 1})^{(1,0)}$ & $3$
& $\sum_{l_i=0}^{J_Y}  Y^{l_1} \,\psi_\alpha  Y^{l_2}\ \psi_{{\beta}} Y^{l_3} \psi_{\dot{\alpha}}Y^{l_4}\psi_{\dot{\beta}}Y^{J_Y-\sum{l_i}} $\\
\hline
\end{tabular}
\end{center}
The superscripts denote charges under SO(2)$_Z \times $SO(2)$_U$
and we have indicated the energy of fluctuations above the giant
graviton which itself has energy $N$.   As in the case of the
string ground state, anomalous dimensions arising from
interactions between the Zs and the other fields do not grow in
the $N \rightarrow \infty$ limit (see the Appendix).   As above,
anomalous dimension is shifted by $(J-1)g_s/\pi$ presumably
matching a one-loop open string effect that was not considered in
\cite{DP}. Anomalous  dimensions do not arise due to interactions
between $\phi^{1,2}$ and Y or between $\psi$ and Y because of the
symmetrization of fields within the string of Ys.

\paragraph{Higher oscillators and string spectrum from $\CN=4$ theory: }  To construct the higher oscillator states of open strings in analogy with~\cite{BMN} we can insert operators representing string fluctuations into the worldsheet represented by the string of Ys in (\ref{opdef}).  (Again we choose not to  normalize these operators here to have a unit two point function.)  A phase depending on the position of insertion into the string of Ys represents the oscillator level.  In effect, the phases reconstruct the Fourier representation of a momentum state on the string worldsheet in position space along the string.   The operators we can insert include those in (\ref{corresp1}) corresponding to the directions in which the open string has Neumann boundary conditions.
For example:
\begin{equation}
a^7_{-n}|G_N;-1,0 \ket \leftrightarrow
\epsilon_{i_1\cdots i_N} \epsilon^{j_1\cdots
j_N}Z^{i_1}_{j_1}\cdots Z^{i_{N-1}}_{j_{N-1}}
\left( \sum_{l=0}^{J_Y} \, Y^l \, \phi^1 Y^{J_Y - l} \cos({{\pi   n l \over J_Y}})\right)^{i_N}_{j_N}
\label{highstate}
\end{equation}
In addition, although there are no zero modes in directions with Dirichlet boundary conditions for the open string,  there are higher oscillator excitations. These will correspond to insertions of
\begin{eqnarray}
\bar{a}^{i} ~~~~ &\leftrightarrow& ~~~~ D_i Y ~~~~ i=1\cdots4  \nonumber \\
\bar{a}^{5,6} ~~~~ &\leftrightarrow& ~~~~ \phi^{5,6}
\end{eqnarray}
with a position dependent phase $\sin({n\pi l \over J_Y})$:
\begin{equation}
a^5_{-n}|G_N;-1,0 \ket \leftrightarrow
\epsilon_{i_1\cdots i_N} \epsilon^{j_1\cdots
j_N}Z^{i_1}_{j_1}\cdots Z^{i_{N-1}}_{j_{N-1}}
\left( \sum_{l=0}^{J_Y} \, Y^l \, \phi^5 Y^{J_Y - l} \sin({{\pi  n l \over J_Y}})\right)^{i_N}_{j_N}
\label{highstate1}
\end{equation}
The higher fermionic oscillators correspond to similar insertions of $\psi_\alpha$ and
$\psi_{\dot{\alpha}}$ with similar phases. Note that the phase ${{\pi  n l \over J_Y}}$ is
half of the phase appearing in the closed string construction of \cite{BMN}. This is
necessary to correctly reproduce the open string spectrum from field theory.

All operators constructed as in (\ref{highstate},\ref{highstate1}) carry charge $J_Y$ under SO(2)$_Y$ and so to compare the string spectrum (\ref{spectrum})  to the field theory we need only compute the conformal dimension of the operator.
Although the interactions between Zs and the operators within the string of Ys continue to be suppressed as for the zero modes (see the Appendix), the presence of phases in (\ref{highstate},\ref{highstate1}) leads to anomalous dimensions that we must compute in order to match the spectrum (\ref{spectrum})~\cite{BMN}.   Below we will work with the example (\ref{highstate}) but an identical story applies to all the other operators.  (We leave out most of the details of the calculation since it is exactly parallel to the work in~\cite{BMN}.)

To start it is useful to expand the energies (\ref{spectrum}) in a power series in $gN/J_Y^2$:
\begin{equation}
\omega_n = (\Delta - J_Y)_n = 1 + {\pi g N n^2 \over 2 J_Y^2} + \cdots
\label{expansion}
\end{equation}
The classical dimension of (\ref{highstate}) is $\tilde\Delta = N
-1 + J_Y + 1 = N + J_Y$.   In the interacting theory anomalous
dimensions will develop.   To study this we have to compute
correlation functions of (\ref{highstate}).   Even in the free
limit, there are many non-planar diagrams in these correlators
which are not suppressed even at large $N$ because the operator
itself has dimension comparable to $N$~\cite{BBNS}.    However,
within any diagram the interactions between the Ys and themselves
is dominated by planar sub-diagrams because $J_Y \sim \sqrt{N}$
and because when $N$ is large nonplanarity only becomes important
when more that $\sim N^{2/3}$ fields are involved~\cite{BBNS}. The
free contractions between Zs and themselves and the Ys and
themselves give rise to the classical dimension of the operators.
Interactions between Ys and Zs  are discussed in the Appendix and
give rise to an anomalous dimension of $(J-1)g_s/\pi$.  As above,
we expect this shift to be related to a one-loop open string
effect in the pp-wave spacetime.

If we introduce an additional operator $O$ within the string of Ys as in (\ref{highstate}), there
will be further interactions between $O$ and Y which we will discuss here.   There are also interactions between $O$ and Z, which, as we show in the Appendix, lead to small corrections that vanish in the large $N$ limit.
The diagrams connecting $O$ and $Y$  arise  because of the 4 point vertex in the $\CN=4$ theory. For the operator (\ref{highstate}) the
relevant vertex is:
\begin{equation}
g^2_{YM} \tr([Y,\phi^1][\bar{Y},\phi^1] \,
\end{equation}
This can lead to diagrams which exchange the position of $\phi$ and Y in the string of Ys in (\ref{highstate}) while leaving the Zs untouched.  Summing all such diagrams is a computation almost
identical to Appendix A of \cite{BMN}. The only difference arises from the different position
dependent phase in relating the higher oscillators to operators as in (\ref{highstate}).
The result is:
\begin{equation}
(\Delta-J_Y)_n=1+{\pi g N n^2 \over 2J_Y^2}
\end{equation}
This correctly reproduces the first order correction to the energy
in (\ref{expansion}). In fact, the full square root in
(\ref{spectrum}) can be straighforwardly reproduced by iterating
the interaction along the lines of Appendix A in \cite{BMN}.
(Again, the interaction between the $Z$s and the other fields will
shift the $1$ to $1 + (J-1)g_s/\pi$ as discussed earlier.)

 In Sec. 2 we discussed how the $G^2$ low energy fluctuations of $G$ coinciding giants can arise from states with mutiple determinants with intertwined indices (see the discussion around (\ref{inter1}) and (\ref{inter2})).    A similar construction in the Penrose limit of G coinciding D3-branes yields strings stretched between each pair of branes.  The spectra of each of these strings is identical and is reproduced as above.  The presence of a Chan-Paton factor labelling the string endpoints is confirmed by point-splitting the location of the determinant operators in Yang-Mills theory.   At low energies these strings must give rise to a new $U(G)$ gauge theory.  Note, however, that only gauge-invariant operators built from this theory will be visible unless the $U(G)$ is broken by separating the branes.  This is related to the observation in Sec. 2 that when the branes coincide, all the $G^2$ operators describing fluctuations of the multiple giants become identical.

\section{Discussion}

In this paper we have shown how open strings can emerge in pure supersymmetric SU(N) Yang-Mills theory as fluctuations around very heavy states with energy of order N, which map onto spherical D-branes in $\ads{5} \times \sph{5}$.    In a large momentum limit we have explicitly constructed the open string worldsheet and its spectrum of oscillations from Yang-Mills theory.  This shows that the large N limit of SU(N) Yang-Mills theory cannot be described as a closed string theory alone -- open strings are also needed.

The emergence of open strings in this way sheds an interesting light on the difficult problem of achieving a background independent formulation of M theory.  $\CN =4$ Yang-Mills theory is a second-quantized theory and is supposed to give us a non-perturbative definition of string theory.   If this is really so all the different backgrounds of string theory, and their holographic duals should somehow emerge from different sectors of Yang-Mills theory.    Our results in this paper point to one interesting way in which this can happen: large charge limits of Yang-Mills theory give rise to new holographic dualities. Consider  SU(N) theory in a state obtained by acting with $G \gg N$ sub-determinants.   This theory is dual to $\ads{5} \times \sph{5}$ with length scale $R_N={(4\pi {\alpha^\prime}^2 g_s N)}^{1/4}$ and $G$ giant gravitons.     After back-reaction the resulting spacetimes are charged black holes of $\ads{5}$ whose singularity can be understood as a condensate of giant gravitons~\cite{MT} (also see~\cite{BN}).   Since the number of D3 brane giants exceeds the number of background branes creating the spacetime, we expect that near the giants, a new $\ads{5} \times \sph{5}$ will emerge with length scale  $R_G ={(4\pi {\alpha^\prime}^2 g_s G)}^{1/4}$ and an SU(G) holographic dual.  Evidence for this was given in \cite{BN} from considerations of black hole entropy.   From the perspective of the dual SU(N) theory, we expect that in a sector with $G \gg N$ subdeterminants fluctuations give rise to a new SU(G) gauge theory in the infrared.  The results of this paper suggest how this happens.  The open strings running between the giant gravitons in AdS space emerge explicitly from the Yang-Mills theory as we have indicated.   In the appropriate limit these strings will give rise to a new SU(G) Yang-Mills theory dual to  the large-charge limit of the charged $\ads{5}$ black holes.

The above discussion and~\cite{BN} suggest that in a sector where SU(N) Yang-Mills theory carries an R-charge $J \gg N$, the convenient holographic description is in terms of operators on the worldvolume of D3-brane giants that are present in the system.  Curiously, from the point of view of the original Yang-Mills dual to $\ads{5}$, the giant worldvolume is some sort of ``internal" direction, as is the entire $\sph{5}$ on which the giant rotates.   If there were a way to see the 3-brane worldvolume explicitly in Yang-Mills theory, just as string worldsheets are explicit in Penrose limits, this intuition could be made precise.   Note that in the pp-wave limit of $\ads{5}\times\sph{5}$~\cite{BMN} , there are two $R^4$ factors, one coming from the AdS space and the other from the sphere.  Interestingly, the pp-wave limit, as a sector of Yang-Mills with R-charge of order $\sqrt{N}$ is  somehow halfway to the point where giant gravitons dominate the physics of AdS.  In effect, as the R-charge increases, the infrared of  the Yang-Mills theory on $\sph{3} \times R$  that is dual to $\ads{5} \times \sph{5}$  interpolates to the holographic dual to pp-waves~\cite{ppholog} and then back to a new Yang-Mills theory now defined on a different sphere.

\vspace{0.2in} {\leftline {\bf Acknowledgments}} We are grateful
to Samir Mathur  and Steve Corley for useful conversations and
communications.  We thank David Berenstein, Bo Feng, Oleg Lunin
for discussions of calculations in the Appendix. A.N. thanks the
Michigan center for theoretical physics for hospitality and the
organizers of the Mathematics and Physics of extra dimensions
workshop.  This work was supported by DOE grant DE-FG02-95ER40893.

\appendix

\section{Protected dimensions in the $N \rightarrow \infty$ limit}
The operator corresponding to the open string ground state in
(\ref{opdef}),
\begin{equation}
\CO= \epsilon_{i_1\cdots i_N} \epsilon^{j_1\cdots
j_N}Z^{i_1}_{j_1}\cdots Z^{i_{N-1}}_{j_{N-1}} (Y^J )^{i_N}_{j_N}
\label{app}
\end{equation}
is not BPS with respect to the $\CN=4$ algebra. As such,  its
dimension is not protected. In particular, it can develop an
anomalous dimension due to an interaction between the Y and Zs. In
this appendix, we will show that such an anomalous dimension does
not grow in the $N \rightarrow \infty$ limit relevant to pp-waves.

The relevant part of $\CN=4$ SYM is
\begin{equation}
S={1 \over 2\pi g_s} \int d^4 x ~\tr\Bigl({1 \over 2}F_{\mu
\nu}F^{\mu \nu}+D_\mu Z D^\mu \overline{Z} +D_\mu Y D^\mu
\overline{Y}
+[Z,\overline{Z}][Y,\overline{Y}]-2[Z,Y][\overline{Z},\overline{Y}]\Bigr)
\end{equation}
Where $\tr([Z,\overline{Z}][Y,\overline{Y}])$ is the D-term
potential, and $-2\tr([Z,Y][\overline{Z},\overline{Y}])$ is the
F-term potential. According to the argument in Appendix B in
\cite{Constable}, The D-term and gluon exchange cancel at one loop
order (this is based on techniques in previous papers
\cite{DHoker, Skiba} ), so we only need to consider the
contributions from F-term.  The scalar propagators are
\begin{equation}
\bra Z_i^j(x) \overline{Z}_k^l(0) \ket = \bra Y_i^j(x)
\overline{Y}_k^l(0) \ket = \delta^l_i \delta^j_k {2 \pi g_s \over
4 \pi^2}{1 \over |x|^2},
\end{equation}

The two point function can be written as the sum of a free and an
one-loop interacting part:
\begin{equation}
\bra \CO(x) \CO ^*(0) \ket = \bra \CO(x) \CO ^*(0) \ket_f+\bra
\CO(x) \CO^* (0) \ket_{i}.
\end{equation}
The free part is straightforward to calculate. The result is:
\begin{equation}
 \bra \CO(x) \CO^* (0) \ket_f  = {(N-1)! ~ N!^2}{1 \over |x|^{2(N+J-1) }}N^{J-1}
 (\frac{2\pi g_s}{4\pi^2})^{N+J-1}  \equiv
{C \over |x|^{2 \Delta}},
 \end{equation}
where $C\equiv {(N-1)! ~ N!^2}N^{J-1}(\frac{2\pi
g_s}{4\pi^2})^{N+J-1} $ and $\Delta=(N+J-1) $. The factor of
$(N-1)!$ counts the number of contractions between the Zs, $N!^2$
is the result of contracting four $\epsilon$ tensors in pairs and
$N^{J-1}$ results from $J-1$ loops from contracting the Ys
planarly.

The interacting part receives contributions from F-term potential
\begin{equation}
V_F=-2\tr([Z,Y][\overline{Z},\overline{Y}])=-2\tr(ZY\overline{Z}\overline{Y}+
YZ\overline{Y}\overline{Z}-ZY\overline{Y}\overline{Z}-YZ\overline{Z}\overline{Y})
\end{equation}
The calculation of the one-loop interactive part is to insert the
operator $-\frac{1}{2\pi g_s}V_F$, then integrate the three point
function over the position of the inserted operator. First we
consider inserting $-\frac{1}{\pi
g_s}\tr(ZY\overline{Y}\overline{Z}(y))$,

\begin{eqnarray*}
\int d^4 y \bra \CO(x) \CO^* (0) (-\frac{1}{\pi
g_s})\tr(ZY\overline{Y}\overline{Z}(y))\ket = {1 \over
|x|^{2\Delta}} \Bigl(- {1 \over{ \pi g_s}}\Bigr)
\Bigl({2 \pi g_s \over 4 \pi^2}\Bigr)^{N+J+1}|x|^4 \int d^4 y ~{1 \over |y|^4|x-y|^4}\\
 \times \epsilon_{i_1\cdots i_N}\epsilon^{j_1\cdots
j_N}\epsilon_{k_1\cdots k_N}\epsilon^{l_1\cdots l_N}\bra
Z^{i_1}_{j_1}\cdots Z^{i_{N-1}}_{j_{N-1}}(Y^J)^{i_N}_{j_N}
\overline{Z}^{k_1}_{l_1}\cdots
\overline{Z}^{k_{N-1}}_{l_{N-1}}(\overline{Y}^J)^{k_N}_{l_N}:\tr(ZY\overline{Y}\overline{Z}):\ket
\end{eqnarray*}
we extract the log divergent piece
\begin{equation}
|x|^4 \int d^4 y ~{1 \over |y|^4|x-y|^4}=4\pi^2\log(|x|\Lambda)
\end{equation}
and the large N power piece
\begin{eqnarray}
&&\epsilon_{i_1\cdots i_N}\epsilon^{j_1\cdots
j_N}\epsilon_{k_1\cdots k_N}\epsilon^{l_1\cdots l_N}\bra
Z^{i_1}_{j_1}\cdots Z^{i_{N-1}}_{j_{N-1}}(Y^J)^{i_N}_{j_N}
\overline{Z}^{k_1}_{l_1}\cdots
\overline{Z}^{k_{N-1}}_{l_{N-1}}(\overline{Y}^J)^{k_N}_{l_N}:Z^a_bY^b_c\overline{Y}^c_d\overline{Z}^d_a:\ket
\nonumber \\ \nonumber &=&(N-1)^2(N-2)!\epsilon_{i_1\cdots
i_N}\epsilon^{j_1\cdots j_N}\epsilon_{k_1\cdots
k_N}\epsilon^{l_1\cdots l_N}\\ \nonumber &\times &
\delta^{i_1}_{l_1}\cdots\delta^{i_{N-2}}_{l_{N-2}}\delta^{k_1}_{j_1}\cdots\delta^{k_{N-2}}_{j_{N-2}}
\delta^{i_{N-1}}_a\delta^d_{j_{N-1}}\delta^{k_{N-1}}_b\delta^a_{l_{N-1}}\bra
(Y^J)^{i_N}_{j_N}(\overline{Y}^J)^{k_N}_{l_N}
:Y^b_c\overline{Y}^c_d: \ket \\ \nonumber
&=&(N-1)^2(N-2)!^3(\delta^a_a\delta^{l_N}_{i_N}-\delta^{l_N}_a\delta^a_{i_N})
(\delta^d_b\delta^{j_N}_{k_N}-\delta_{k_N}^d\delta_b^{j_N}) \bra
(Y^J)^{i_N}_{j_N}(\overline{Y}^J)^{k_N}_{l_N}
:Y^b_c\overline{Y}^c_d: \ket
\\ \nonumber
&=&(N-1)^3(N-2)!^3 (\bra\tr(Y^J\overline{Y}^J)
:\tr(Y\overline{Y}): \ket-\bra\tr(Y^J:Y\overline{Y}:
\overline{Y}^J)\ket) \\  &=&(N-1)^3(N-2)!^3(JN^{J+1})
\end{eqnarray}
where in the last step the first term dominate and gives
$JN^{J+1}$ (The factor of $J$ comes from $J$ ways of contracting).
So
\begin{eqnarray} \label{9}
\int d^4 y \bra \CO(x) \CO ^*(0) (-\frac{1}{\pi
g_s})\tr(ZY\overline{Y}\overline{Z}(y))\ket
=-\frac{Jg_s}{\pi}\frac{C}{|x|^{2\Delta}}\log(|x|\Lambda)
\end{eqnarray}
Similarly
\begin{eqnarray} \label{10}
\int d^4 y \bra \CO(x) \CO^* (0) (-\frac{1}{\pi
g_s})\tr(YZ\overline{Z}\overline{Y}(y))\ket
=-\frac{Jg_s}{\pi}\frac{C}{|x|^{2\Delta}}\log(|x|\Lambda)
\end{eqnarray}
Now we consider inserting $\frac{1}{\pi
g_s}\tr(ZY\overline{Z}\overline{Y}(y))$,
\begin{eqnarray*}
\int d^4 y \bra \CO(x) \CO^* (0) (\frac{1}{\pi
g_s})\tr(ZY\overline{Z}\overline{Y}(y))\ket = {1 \over
|x|^{2\Delta}} \Bigl( {1 \over{ \pi g_s}}\Bigr)
\Bigl({2 \pi g_s \over 4 \pi^2}\Bigr)^{N+J+1}|x|^4 \int d^4 y ~{1 \over |y|^4|x-y|^4}\\
 \times \epsilon_{i_1\cdots i_N}\epsilon^{j_1\cdots
j_N}\epsilon_{k_1\cdots k_N}\epsilon^{l_1\cdots l_N}\bra
Z^{i_1}_{j_1}\cdots Z^{i_{N-1}}_{j_{N-1}}(Y^J)^{i_N}_{j_N}
\overline{Z}^{k_1}_{l_1}\cdots
\overline{Z}^{k_{N-1}}_{l_{N-1}}(\overline{Y}^J)^{k_N}_{l_N}:\tr(ZY\overline{Z}\overline{Y}):\ket
\end{eqnarray*}
where the large N power piece is
\begin{eqnarray}
&&\epsilon_{i_1\cdots i_N}\epsilon^{j_1\cdots
j_N}\epsilon_{k_1\cdots k_N}\epsilon^{l_1\cdots l_N}\bra
Z^{i_1}_{j_1}\cdots Z^{i_{N-1}}_{j_{N-1}}(Y^J)^{i_N}_{j_N}
\overline{Z}^{k_1}_{l_1}\cdots
\overline{Z}^{k_{N-1}}_{l_{N-1}}(\overline{Y}^J)^{k_N}_{l_N}:Z^a_bY^b_c\overline{Z}^c_d\overline{Y}^d_a:\ket
\nonumber \\ \nonumber &=&(N-1)^2(N-2)!\epsilon_{i_1\cdots
i_N}\epsilon^{j_1\cdots j_N}\epsilon_{k_1\cdots
k_N}\epsilon^{l_1\cdots l_N}\\ \nonumber &\times &
\delta^{i_1}_{l_1}\cdots\delta^{i_{N-2}}_{l_{N-2}}\delta^{k_1}_{j_1}\cdots\delta^{k_{N-2}}_{j_{N-2}}
\delta^{k_{N-1}}_b\delta^a_{l_{N-1}}\delta^{i_{N-1}}_d\delta^c_{j_{N-1}}\bra
(Y^J)^{i_N}_{j_N}(\overline{Y}^J)^{k_N}_{l_N}
:Y^b_c\overline{Y}^d_a: \ket \\ \nonumber
&=&(N-1)^2(N-2)!^3(\delta^a_d\delta^{l_N}_{i_N}-\delta^{l_N}_d\delta^a_{i_N})
(\delta^c_b\delta^{j_N}_{k_N}-\delta_{k_N}^c\delta_b^{j_N}) \bra
(Y^J)^{i_N}_{j_N}(\overline{Y}^J)^{k_N}_{l_N}
:Y^b_c\overline{Y}^d_a: \ket
\\ \nonumber
&=&(N-1)^2(N-2)!^3 \bra\tr(Y^J Y\overline{Y}^J \overline{Y}) \ket
\\  &=&(N-1)^2(N-2)!^3(N^{J+2})
\end{eqnarray}
So
\begin{eqnarray} \label{12}
\int d^4 y \bra \CO(x) \CO ^*(0) (\frac{1}{\pi
g_s})\tr(ZY\overline{Z}\overline{Y}(y))\ket
=\frac{g_s}{\pi}\frac{C}{|x|^{2\Delta}}\log(|x|\Lambda)
\end{eqnarray}
Similarly
\begin{eqnarray} \label{13}
\int d^4 y \bra \CO(x) \CO ^*(0) (\frac{1}{\pi
g_s})\tr(YZ\overline{Y}\overline{Z}(y))\ket
=\frac{g_s}{\pi}\frac{C}{|x|^{2\Delta}}\log(|x|\Lambda)
\end{eqnarray}
So the total contribution by summing
(\ref{9})(\ref{10})(\ref{12})(\ref{13}) is
\begin{equation}
\bra \CO(x) \CO^* (0)
\ket_{i}==-\frac{2(J-1)g_s}{\pi}\frac{C}{|x|^{2\Delta}}\log|x|\Lambda
\end{equation}
So the anomalous dimension is $(J-1)\frac{g_s}{\pi}$. It is
interesting to note when $J=1$, the anomalous dimension cancels.
It is what we should expect, since when $J=1$ this operator is
half BPS operator representing giant graviton in a slightly
different direction.

\paragraph{Anomalous dimensions for operators corresponding to small
fluctuations} In Section 2, we proposed operators
(\ref{mode3},\ref{mode1},\ref{mode2}) corresponding to small
fluctuations of spherical D3 branes, and mapped the dimension of
these operators to the excitation spectrum obtained in \cite{DJM}.
Being non-BPS, these operators can develop anomalous dimensions. A
calculation almost identical to the one above reveals that the
anomalous dimensions of these operators are given by the above
$(J-1)\frac{g_s}{\pi}$ plus terms that are at most of order ${g_s
\over N^2}$ and are therefore suppressed in the large $N$ limit.

\end{document}